\documentclass[prX,showpacs,amsmath,amssymb]{revtex4}

\usepackage[latin1]{inputenc}
\usepackage{graphicx}
\providecommand{\prt}[1]{\left( #1 \right)}
\providecommand{\quotes[1]}{``#1"}
\newcommand{\bracket}[1]{\langle #1\rangle}
\providecommand{\erdren}{Erd\H{o}s-R\'enyi }

\begin{document}

\title{Distance distribution in random graphs and
application to networks exploration}

\author{Vincent D. Blondel}
\author{Jean-Loup Guillaume}
\author{Julien M. Hendrickx}
\author{Raphaël M. Jungers}
\email{{vincent.blondel,jean-loup.guillaume,julien.hendrickx,raphael.jungers}@uclouvain.be}
\thanks{The research reported here was
partially supported by the ``Communaut\'e francaise de Belgique -
Actions de Recherche Concert\'ees", by the EU HYCON Network of
Excellence (contract number FP6-IST-511368), and by the Belgian
Programme on Interuniversity Attraction Poles initiated by the
Belgian Federal Science Policy Office.
The scientific responsibility rests with its authors. Julien Hendrickx and Rapha\"el
Jungers are FNRS fellows (Belgian Fund for Scientific Research).}

\affiliation{Department of Mathematical Engineering,
Universit\'e catholique de Louvain, 4 avenue Georges Lemaitre,
B-1348 Louvain-la-Neuve, Belgium}


\begin{abstract}
We consider the problem of determining the proportion of edges
that are discovered in an \erdren graph when one constructs all
shortest paths from a given source node to all other nodes. This
problem is equivalent to the one of determining the proportion of
edges connecting nodes that are at identical distance from the
source node.  The evolution of this quantity with the probability
of existence of the edges exhibits intriguing oscillatory
behavior. In order to perform our analysis, we introduce a new way
of computing the distribution of distances between nodes. Our
method outperforms previous similar analyses and leads to
estimates that coincide remarkably well with numerical
simulations. It allows us to characterize the phase transitions
appearing when the connectivity probability varies.
\end{abstract}

\maketitle

\section{Introduction}\label{section-intro}

The small-world phenomenon  has attracted increasing attention
over the last few years \cite{rand-graphs,milgram}. In a
small-world network, the average distance between two nodes is
small as compared to the total number of nodes. In many natural
networks, it is typically of the order of $\log(n)$ ($n$ is the
total number of nodes) and several models have been proposed to
explain this  phenomenon (see, e.g. \cite{rand-graphs,
doro2003evolution, watt_sw}). In some applications though, one is
interested not only in this so-called \quotes{average inter-vertex
distance}, but in the whole inter-vertex distance distribution.

Even though this {distribution} is {of much interest,} it has not
been studied very much in the literature. A theoretical method for
the computation of the distances in uncorrelated random networks
of infinite size has been proposed by Dorogovtsev et al. in 2003
\cite{doro_dist}. In $2004, $ Fronczak et al. have analyzed the
distance between nodes for a wide class of random networks of
finite size that generalizes the \erdren graphs, the so-called
uncorrelated random networks with hidden variables {\cite{fron}}.
They propose an approximation of the distribution of the distance
between nodes that performs well for a certain range of the
parameter values. Their formula has the advantage of being simple
and analytical, but the approximations done in the calculations
lead to significative differences with
the numerical evidence for some ranges of the parameters.\\

Our work is motivated by the analysis of algorithms that have been
recently developped for analysing networks, such as the internet.
A typical way of doing that is to use the freeware
\texttt{traceroute}, that provides the user a short path from his
computer to any other one in the internet.  In the ASP model (All
Shortest Paths), introduced to model this strategy, one chooses a
particular node $s$ of the network, and then constructs all
shortest paths from $s$ to all other nodes of the network
\cite{jlg}. Some edges of the network may not belong to any of
these shortest paths and so they are left undiscovered. The
problem considered in \cite{jlg} is that of determining the
proportion of edges of the network that are discovered. Thus the
question is: ``what is the proportion of edges that are on {at
least one shortest path starting from the source}?''. As pointed
out in \cite{jlg}, the edges that are {not discovered are exactly
those} connecting nodes that are at identical distances from the
source. Indeed, if an edge connects two equidistant nodes, it
cannot be on a shortest path from the source, since any path using
this edge (say going from $v_1$ to $v_2$) can be shortened by
going directly to $v_2$ via the shortest path to $v_2.$
{Conversely}, if an edge links two nodes that are not at the same
distance, then it links a node $v_1$ at a certain distance $d$ to
a node $v_2$ at a distance $d+1,$ and {at least one shortest} path
to $v_2$ passes through this edge. We are therefore interested in
computing the number of edges connecting nodes that are at the
same distance from the source. Other models exist for representing
network analysis strategies. For instance, \cite{jlg} introduces
the USP model (for Unique Shortest Path). In the USP model one
chooses only one shortest path from the source to each node in the
graph and so there are possibly more edges that are left
undiscovered. Our work is also relevant to the analysis of this
model, as it counts the proportion of edges that are never found
by any single or multiple USP searches.

\begin{figure}
\centering
\includegraphics[scale = 0.5]{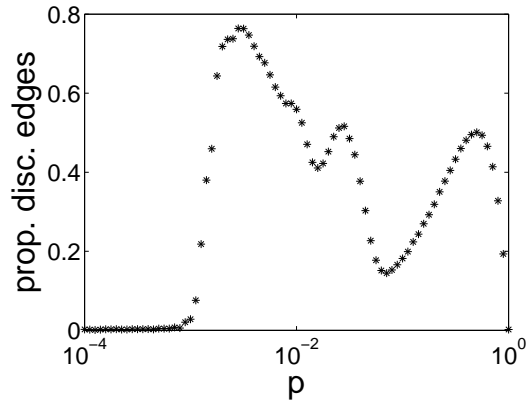}
\caption{Evolution with $p$ (log-scale) of the proportion of edges
that lie on a shortest path in an \erdren random graph with
$n=1000$ vertices. Each value is computed by averaging the
observations made on 1000 graphs \cite{jlg}.}\label{fig-camel}
\end{figure}

In \cite{jlg}, massive numerical simulations have been {performed}
to analyze the proportion of edges that are on shortest paths in
\erdren graphs.  In such random graphs, edges are all equally
likely to be present and the probability of presence is given by
some fixed probability $p$. We do not consider self loops nor
multiple edges. So, for constructing an \erdren graph, one needs
to fix two parameters: the number of nodes $n$ and the probability
of existence for every edge $p$. {As shown in FIG.
\ref{fig-camel}, the proportion of edges that are discovered in
the ASP model presents an interesting dependence in the parameter
$p$}. One can directly explain some characteristics of this curve.
When $p$ is very small the graph is highly disconnected and
consists in small connected components. Most edges do therefore
not belong to any path starting from the source, and the
proportion of observed edges is close to zero. Conversely, if $p$
is very high, the graph is almost complete, and every shortest
path has length one.  So $n-1$ edges are found, while there are
almost $\frac{1}{2}n(n-1)$ edges in the graph, and
thus the proportion also vanishes.\\

The aims of this paper are first to introduce a new simple model
of inter-vertex distances in \erdren graphs that can be used to
compute the curve of FIG. \ref{fig-camel} without any numerical
experiment, and second to analyze the oscillating behavior of this
curve {and explain the phase transitions appearing with variations
of the graph connectivity.} {Note that similar oscillating
behaviors in random graphs have recently} been observed
\cite{holyst-oscillant}, and that these phenomena seem to open
challenging questions in random graphs theory. {{This paper
proposes a precise analysis of such an oscillating behavior} in
the simple theoretical framework of \erdren graphs. {One could
imagine} exploiting these oscillations to optimize the design of a
network or to develop method for its analysis, although this is
beyond the scope of this paper. Besides, such applications of the
concepts developed here would probably require some further
analysis {and extension of our results}, because real networks
often exhibit non-trivial correlations between nodes that do not
occur in \erdren graphs. These extensions would however most
likely not lead to the derivation of simple analytical solutions
providing an intuitive understanding of the
phenomena as it is done here.}\\

The remainder of the paper is organized as follows. In Section
\ref{section-modelisation}, we introduce a recurrence equation
allowing to evaluate the inter-vertex distance distribution for
\erdren graphs, and compare to previously published results
\cite{fron}. From this function we derive a theoretical expression
for values shown on FIG. \ref{fig-camel}. In Section
\ref{section-analysis} we analyze this curve, we characterize the
phase transitions, and give analytical expressions in different
phases (proved in Appendix \ref{section-ap}). In Section
\ref{section-conclusions} we conclude and make some remarks on
practical applications of the phenomena studied in the paper.

\section{Approximation of intervertex distance distribution}\label{section-modelisation}

In this section we propose an approximation for inter-vertex
distance distribution in \erdren graphs. We compare our results to
those obtained by Fronczak et al. \cite{fron} in a more generic
situation, and show how our results outperform theirs in the
particular case of \erdren graphs. We also analyze the accuracy of
our model and its dependence on the graph connectedness. We then
use our inter-vertex distance distribution to estimate the
proportion of equidistant pairs of nodes.\\

In the sequel, we consider the distance between a randomly
selected node and a fixed but initially randomly selected
\quotes{source node}. Since this source is randomly selected, all
results obtained for the distance probability can also be applied
to the distance between two randomly selected vertices. Let $F_d$
be the probability for a randomly selected node to be at a
distance larger than $d$ from the source, that is, the probability
that there is no path of length smaller than or equal to $d$ from
the source to this node. The probability $f_d$ for the node to be
at a distance exactly $d$ of the source is then given by $f_d
=F_{d-1}- F_d$. Obviously, $F_0 = 1-\frac{1}{n}$. We now derive a
recurrence relation allowing the computation of $F_d$ for higher
values of $d$. A node is at a distance larger than $d$ from the
source if it is not the source itself, which happens with
probability $1-\frac{1}{n}$, and if it is connected to no node at
distance less than $d$ from the source, which happens with
probability $(1-p)^{n_d},$ $n_d$ being the number of nodes at
distance less than $d$ from the source. We have therefore the
following simple relation:
\begin{equation}\label{eq-rec}
F_d=\prt{1-\frac{1}{n}}\sum_{k=1}^{n-1}{P[n_d=k](1-p)^{k}},
\end{equation}
where $P[n_d=k]$ denotes the probability that $n_d=k$. In order to
express the probability $F_d$, we should thus know the
distribution of $n_d$. We approximate this quantity to be always
exactly equal to its expectation $\bracket{n_d}=\prt{1-F_{d-1}}n$.
Introducing this approximation in (\ref{eq-rec}) we obtain a
recurrence relation for $F.$
\begin{equation}\label{eq-rec-app}
F_d=\prt{1-\frac{1}{n}}(1-p)^{(1-F_{d-1})n},
\end{equation}
which allows us to compute $f_d$ for any $d$.  This formula is
different, but provably equivalent to Equation $(6)$ in
\cite{baronchelli06} that has been derived independently for other
purposes.\\

In \cite{fron}, Fronczak et al. propose an expression for the
intervertex distance distribution of any \quotes{random graph with
hidden variables}, that are generalizations of \erdren graphs. In
these graphs, two nodes $i$ and $j$ are connected with a
probability $p_{i,j}=h_ih_j/\beta$, where each node $v$ has its
own ``hidden variable'' $h_v,$ and $\beta = \bracket{h}n.$ So, in
a large graph, the hidden variables represent the expected degree
of the vertices.  In the particular case of \erdren graphs, that
is when $h_v = np$ for all $v\in V$, their expression for the
function $F$ of inter-vertex distance distribution reduces to:
\begin{equation}\label{eq-fron}
F_d= e^{-\frac{1}{n}(np)^d}.
\end{equation}
This result has a straightforward interpretation as the solution
of an other recurrence equation on $d$, although it is not
obtained in that way in \cite{fron}. A vertex $i$ is at a distance
larger than $d$ from the source node if all its neighbors are at
distance larger than $d-1$ from the source. Approximating the
number of neighbors by its expectation $np$ and neglecting the
dependence effects, one obtains the recurrence $F_d =
F_{d-1}^{np}$. The relation (\ref{eq-fron}) is then re-obtained by
taking $F_0 = e^{-\frac{1}{n}} \simeq 1-\frac{1}{n}$ as initial
condition. Numerical experiments confirm indeed that taking
$e^{-\frac{1}{n}}$ or $ 1-\frac{1}{n}$ as initial condition has
no influence on the results if $n$ is sufficiently large.\\

\begin{figure}
\centering
\begin{tabular}{cc}
\includegraphics[scale = 0.3]{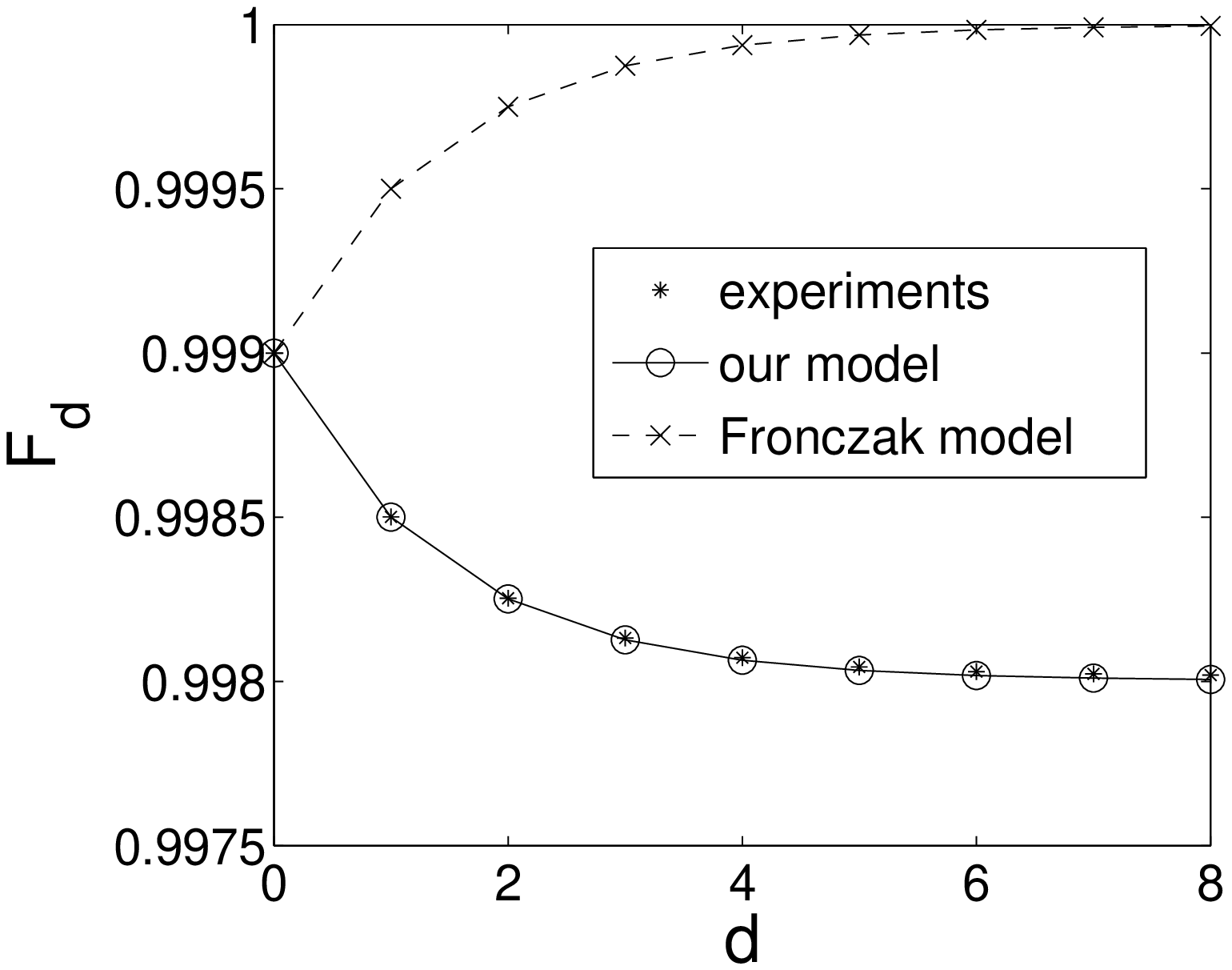}&
\includegraphics[scale = 0.3]{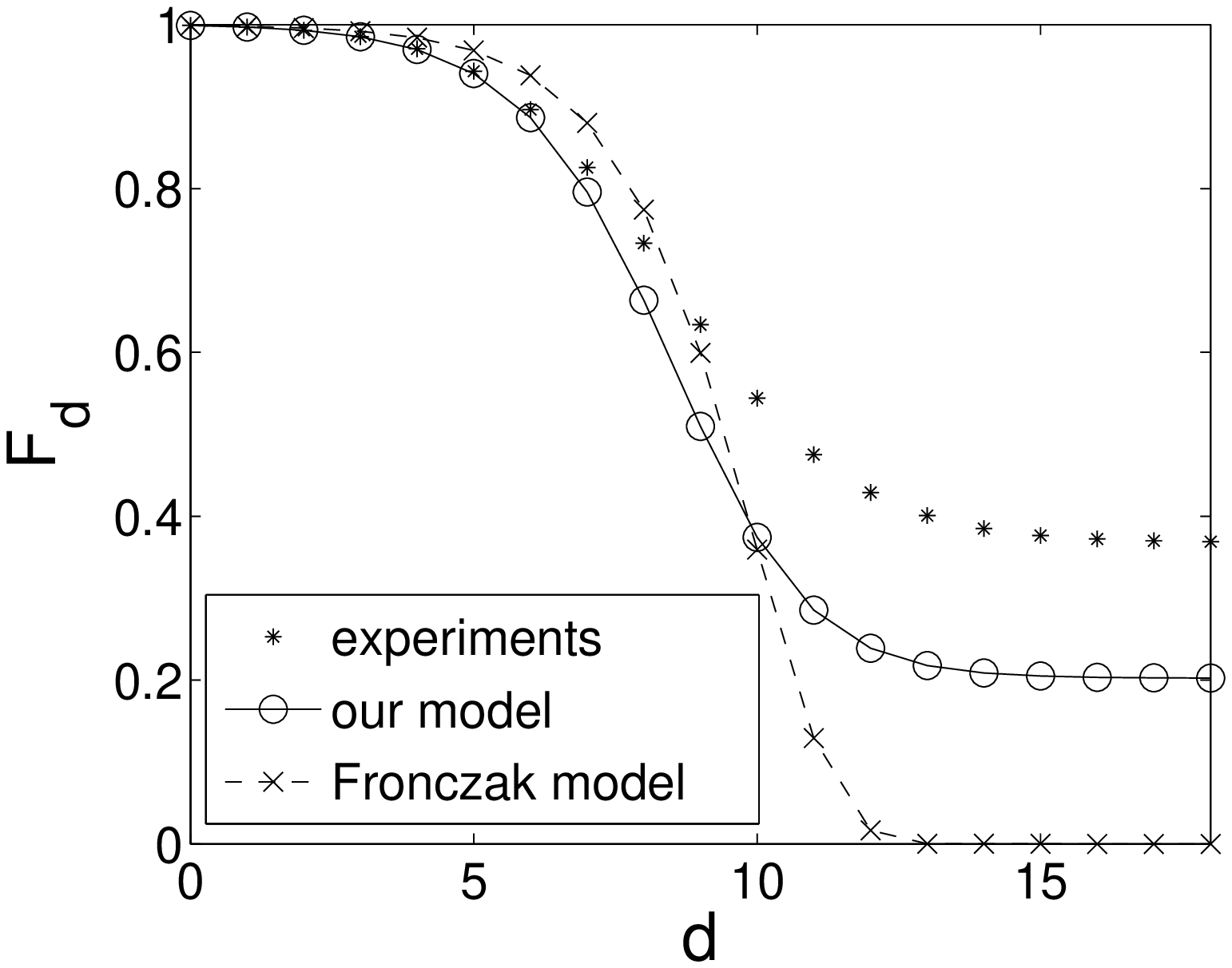} \\(a)&(b)\\ \\

\includegraphics[scale = 0.3]{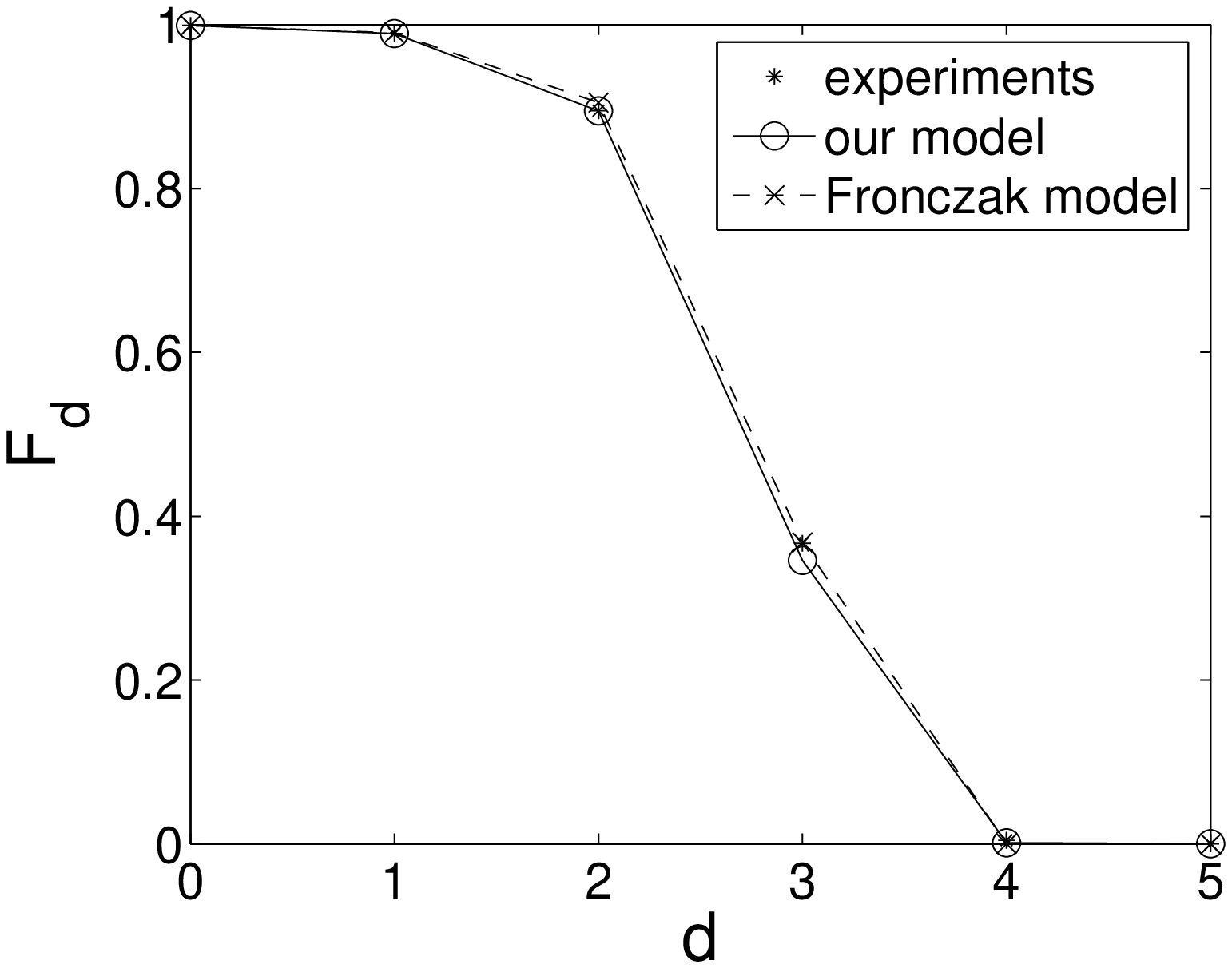}&
\includegraphics[scale = 0.3]{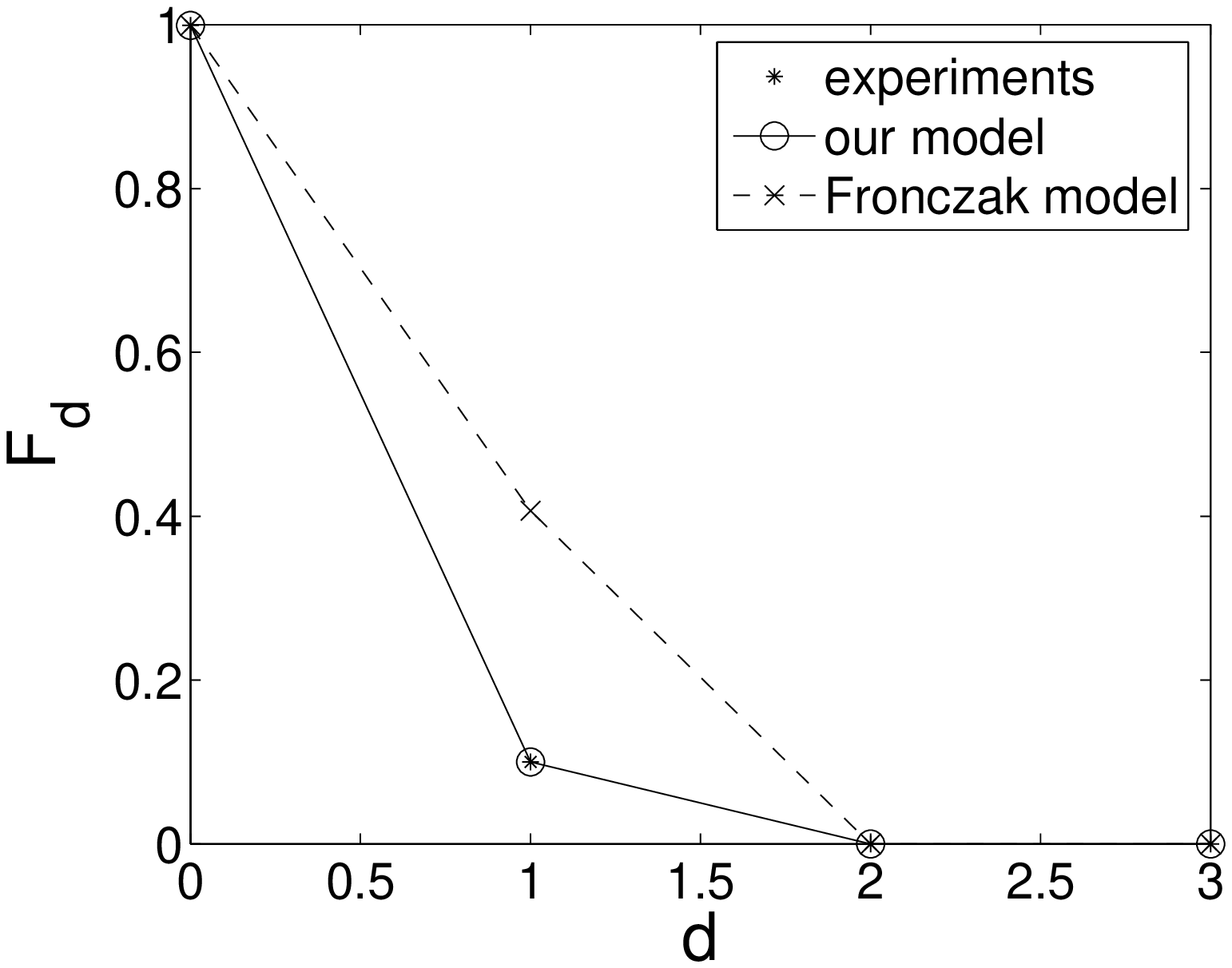}\\(c) & (d)
\end{tabular}

\caption{Evolution of $F_d$, the probability for a random node to
be at a distance larger than $d$ from the source node, for
$n=1000$ nodes and for (a) $np = 0.5$, (b) $np = 2$, (c) $np =
10$, (d) $np = 900$. The three curves represent the experimental
observations (averaged on 500 graphs), our model, and the model of
Fronczak et al given in
\cite{fron}.\label{fig-comp_F(d)}}\end{figure}

In FIG. \ref{fig-comp_F(d)} we compare the predictions from the
two models, with numerical results. One can see that both models
perform very well when the average degree $np$ is significantly
larger than $1$ and if $p$ is not too big, as in FIG.
\ref{fig-comp_F(d)}(c). For an average degree $np<1$, that is
below the emergence of the giant connected component (see
\cite{rand-graphs}), our results match approximately the
experimental observations while Fronczak et al.'s model is not
valid as it gives an increasing curve (see FIG.
\ref{fig-comp_F(d)}(a)). For values of $np$ larger than but close
to 1, both models present significant errors but ours is closer to
the experimental observations (see FIG. \ref{fig-comp_F(d)}(b)).
Finally, for a large $p$, one can see in FIG.
\ref{fig-comp_F(d)}(d) that our results match the experimental
data very well while those obtained with the model of
\cite{fron} are significantly different.\\

The fact that the model derived in \cite{fron} behaves very
differently from our model for a certain range of values of $p$
may seem surprising. Our derivation presents indeed various
similarities with the interpretation of Fronczak et al.'s model as
a solution of a recursive equation. Three reasons can however
explain why a model based on this interpretation gives less
accurate results than ours. First, for $np<<1,$ the possibility
for the randomly selected node to be the source could not be
neglected, as very few nodes are in the connected component of the
source. When $np$ is larger than but close to $1$, the
approximation that a node has exactly $np$ neighbors leads to
proportionally more important errors. This problem could be solved
by considering a binomial distribution for the number of neighbors
in our interpretation of Fronczak et al.'s model. Finally, for
large values of $p,$ the number of neighbors of the randomly
selected node is large, so that some independence problems are not
negligible. Indeed, the probabilities for two neighbors of the
randomly selected node to be the source are not independent, as
there is exactly one source in the graph.\\

\begin{figure}
\centering

\begin{tabular}{c}
\begin{tabular}{cc}
\includegraphics[scale = .3]{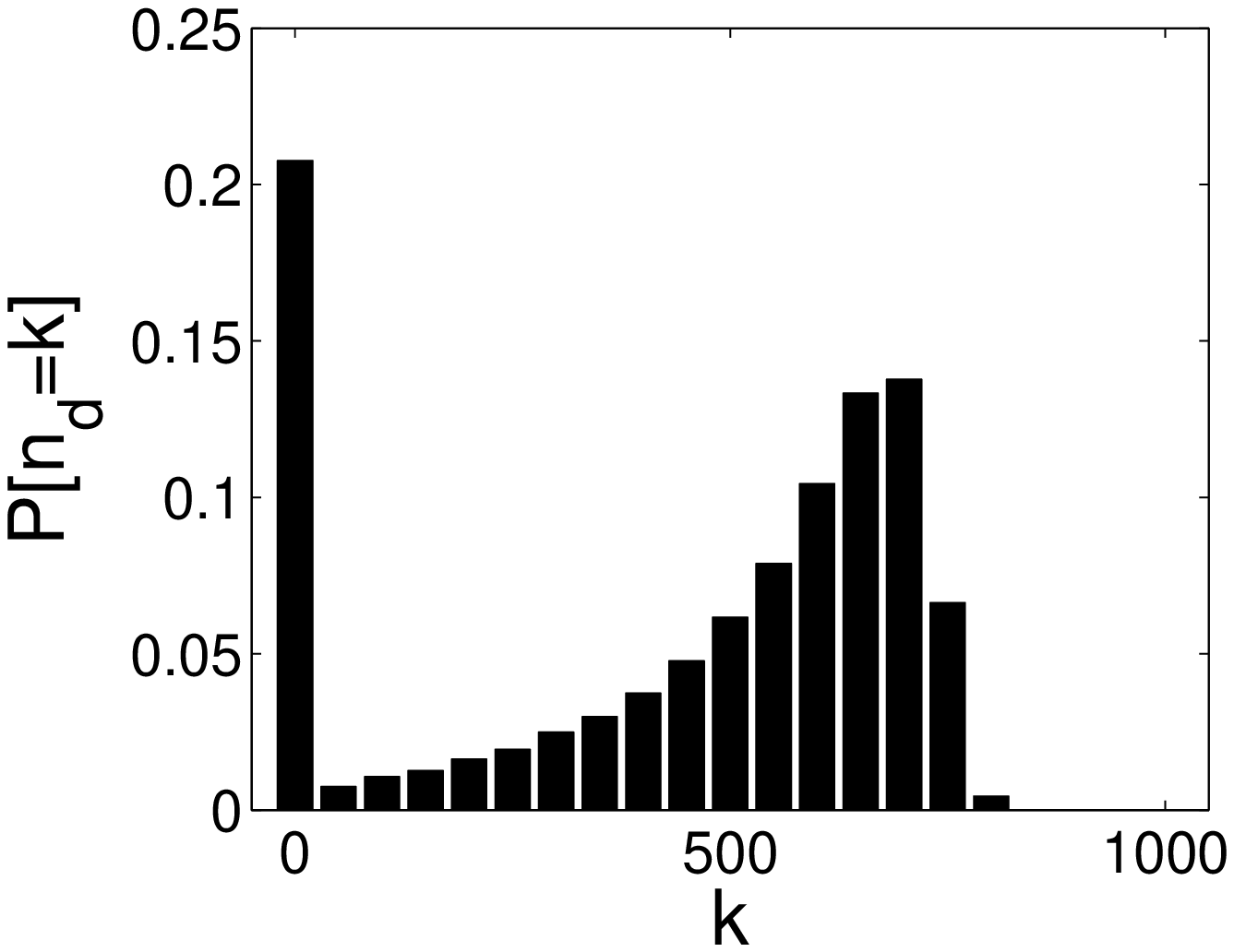} & \includegraphics[scale =
.3]{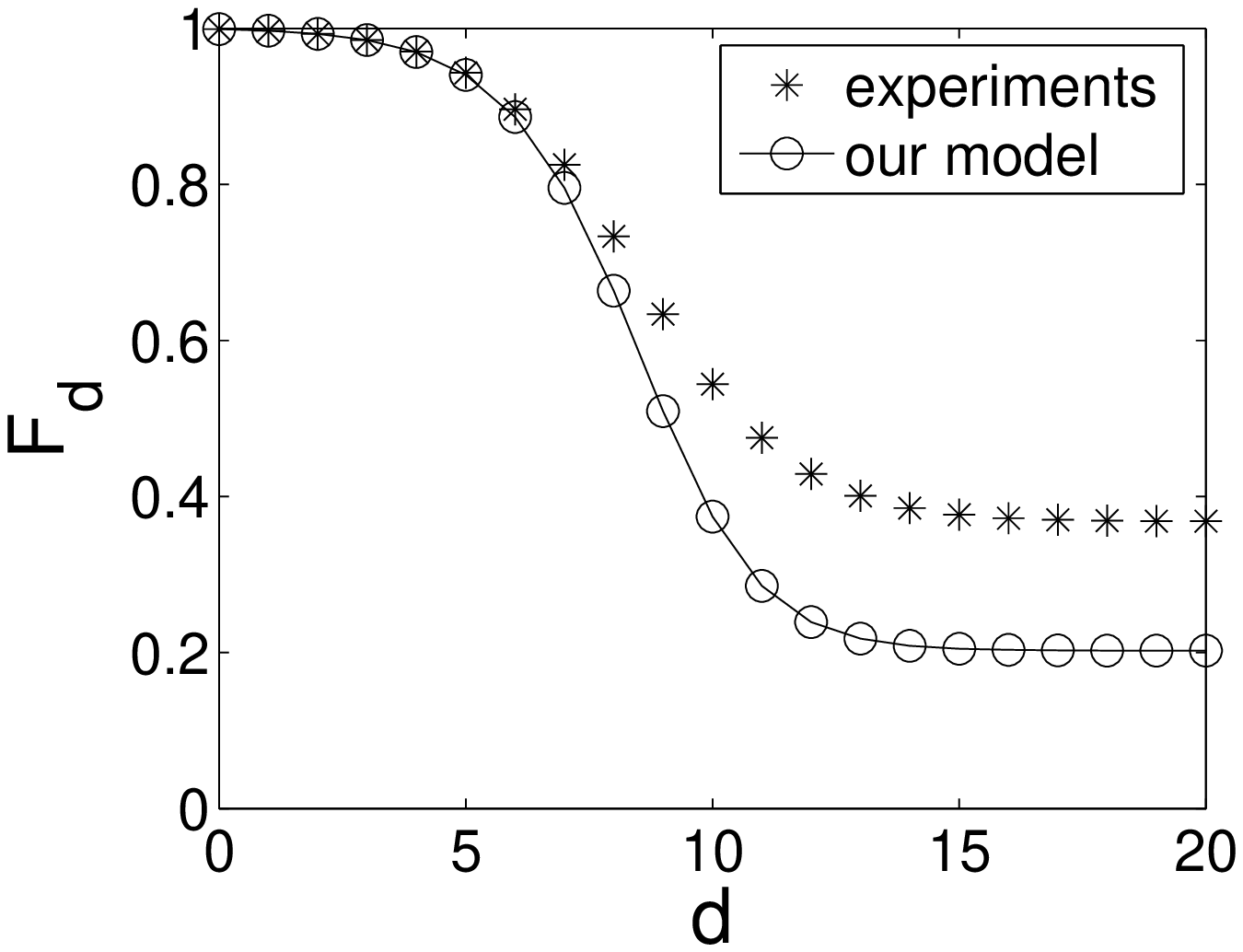}
\end{tabular}\\
(a)\\
\begin{tabular}{cc}
\includegraphics[scale = .3]{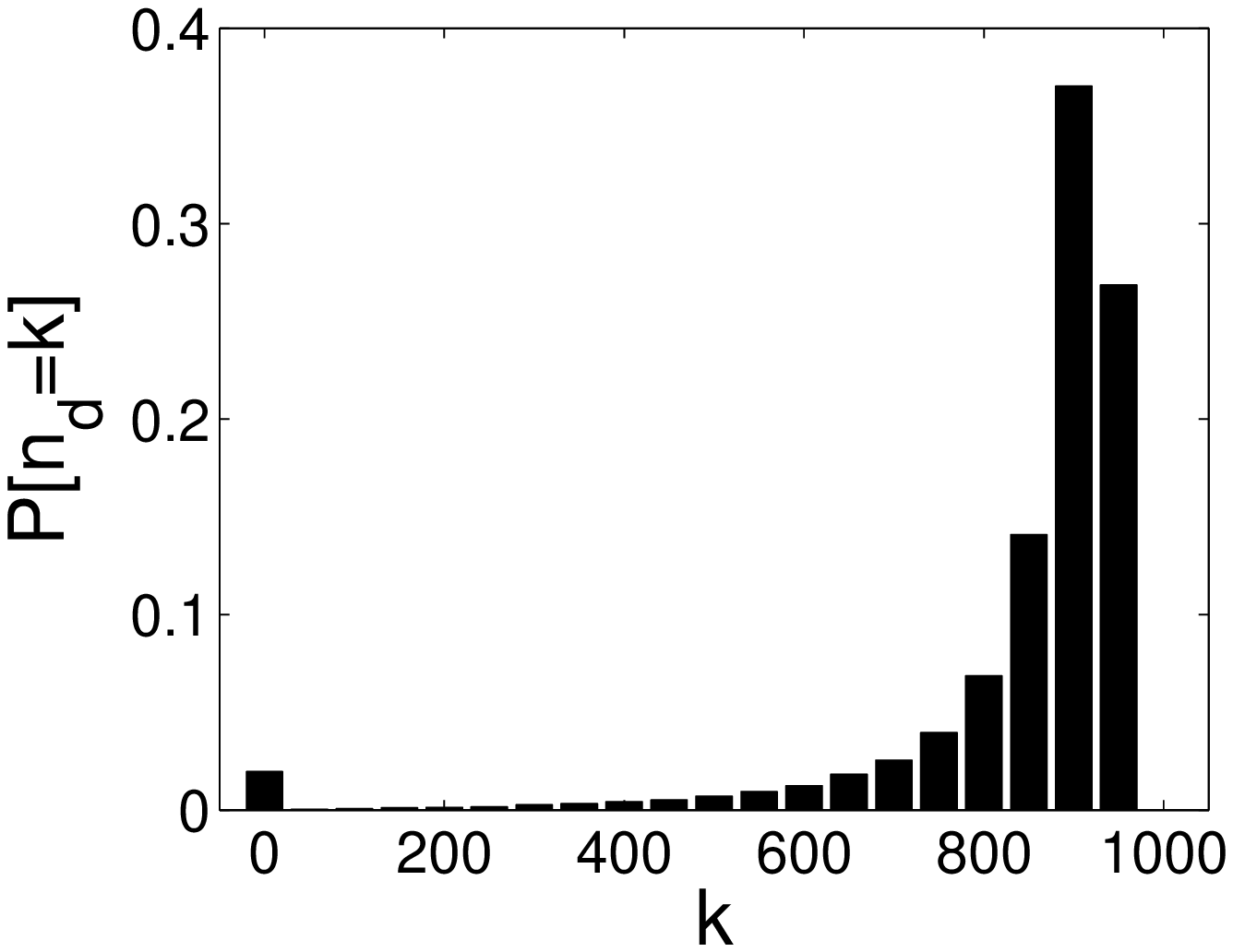} & \includegraphics[scale =
.3]{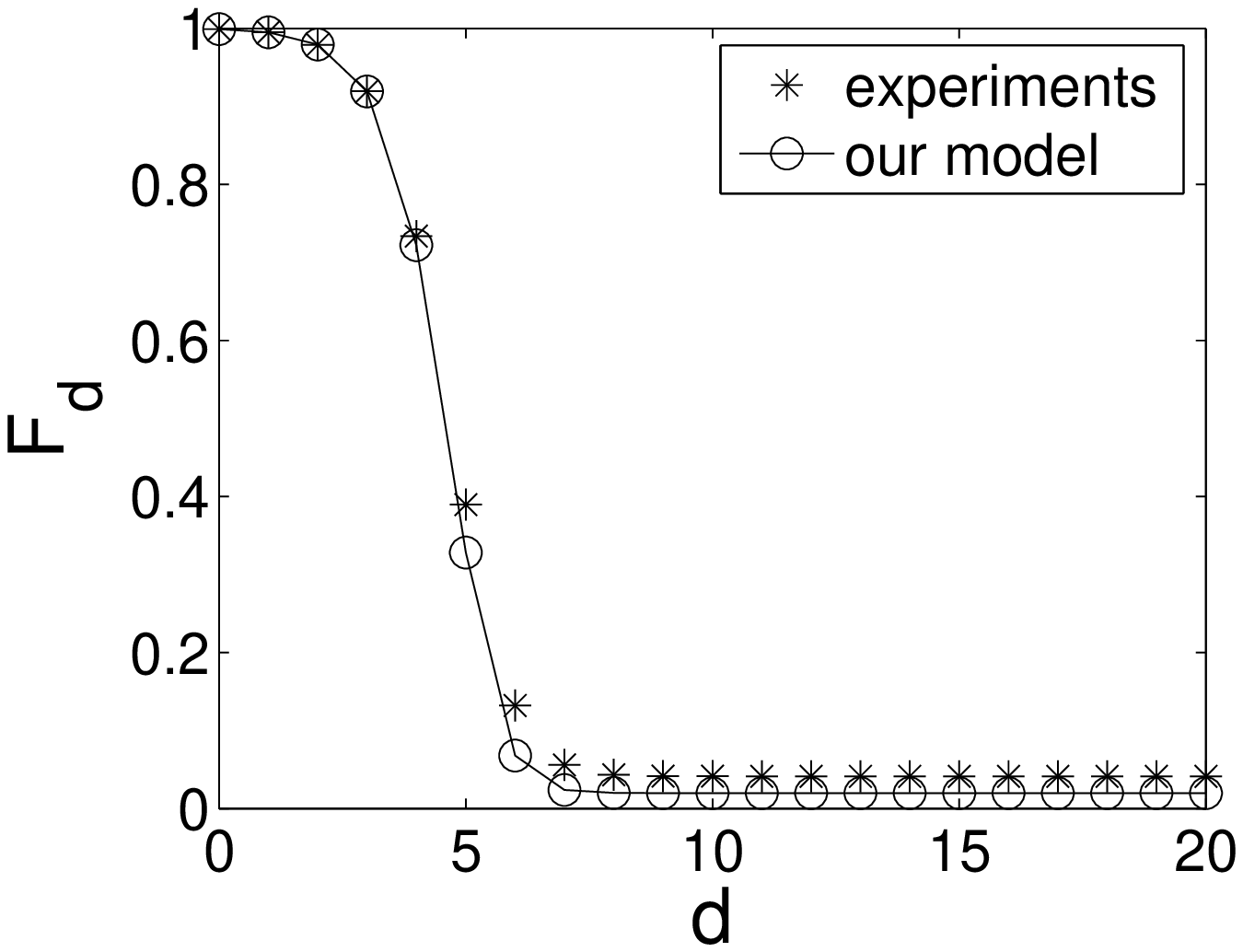}
\end{tabular}\\(b)
\end{tabular}

\caption{{Representation of $P[n_d=k]$, probability that there are
exactly $k$ nodes at distance less than $d$ from the source,
obtained experimentally, in front of $F_d$, proportion of nodes at
a distance larger than $d$ from the source, obtained
experimentally and with our model, for $np=2$ (a) and $np=4$ (b),
with $n=1000$ in both cases. $P[n_d=k]$ is represented for $d=11$
in (a) and for $d=7$ in (b) as typical path lengths are different
when $np=2$ or $np=4$. The distribution in (a) is bimodal as it
contains a large peak around 0, while the peak in (b) is much
smaller. Our approximation of $n_d$ by its average value
$n(1-F_{d-1})$ leads thus to larger errors for $np=2$ (a) than for
$np=4$ (b).} } \label{fig-distri_P(nd=k)}
\end{figure}

The errors of our model, observed for values of $np$ larger than
but close to 1 are due to the approximation mentioned above: To
obtain the recurrence equation (\ref{eq-rec-app}) from
(\ref{eq-rec}), we suppose that the number $n_d$ of vertices at a
distance smaller than $d$ from the source is exactly equal to its
expectancy $n(1-F_{d-1})$ instead of considering its probability
distribution. In this range of parameters, the distribution is far
from being centered because of the existence of a peak around $0$
(see FIG. \ref{fig-distri_P(nd=k)}(a)). For these values indeed,
the graph is not totally connected. If the source happens not to
be in the giant connected component, almost all nodes are at an
infinite distance of it, so that $n_d$ is close to $0$ for any
$d$. The weight of the peak represents thus the probability for a
randomly selected source not to be in the giant connected
component. It is known that when $np$ grows, this probability
tends exponentially to $0$ independently of $n$ \cite[Theorem
5.4]{janson00random}. This problem does therefore only appear when
the average degree $np$ is very small (but larger than $1$),
independently of the size $n$ of the graph. FIG.
\ref{fig-distri_P(nd=k)}(b) shows that the problem is already
almost negligible when $np=4$ (for these values, the giant
connected component already contains more than $98\%$ of the
vertices). Note that for $np<1$ the graph is highly disconnected
so that almost no nodes are at a finite distance from the source.
The distribution $P[n_d=k]$ consists thus only in one peak around
$0$ and is therefore centered.\\

\begin{figure}
\centering
\begin{tabular}{c}
\includegraphics[scale = 0.5]{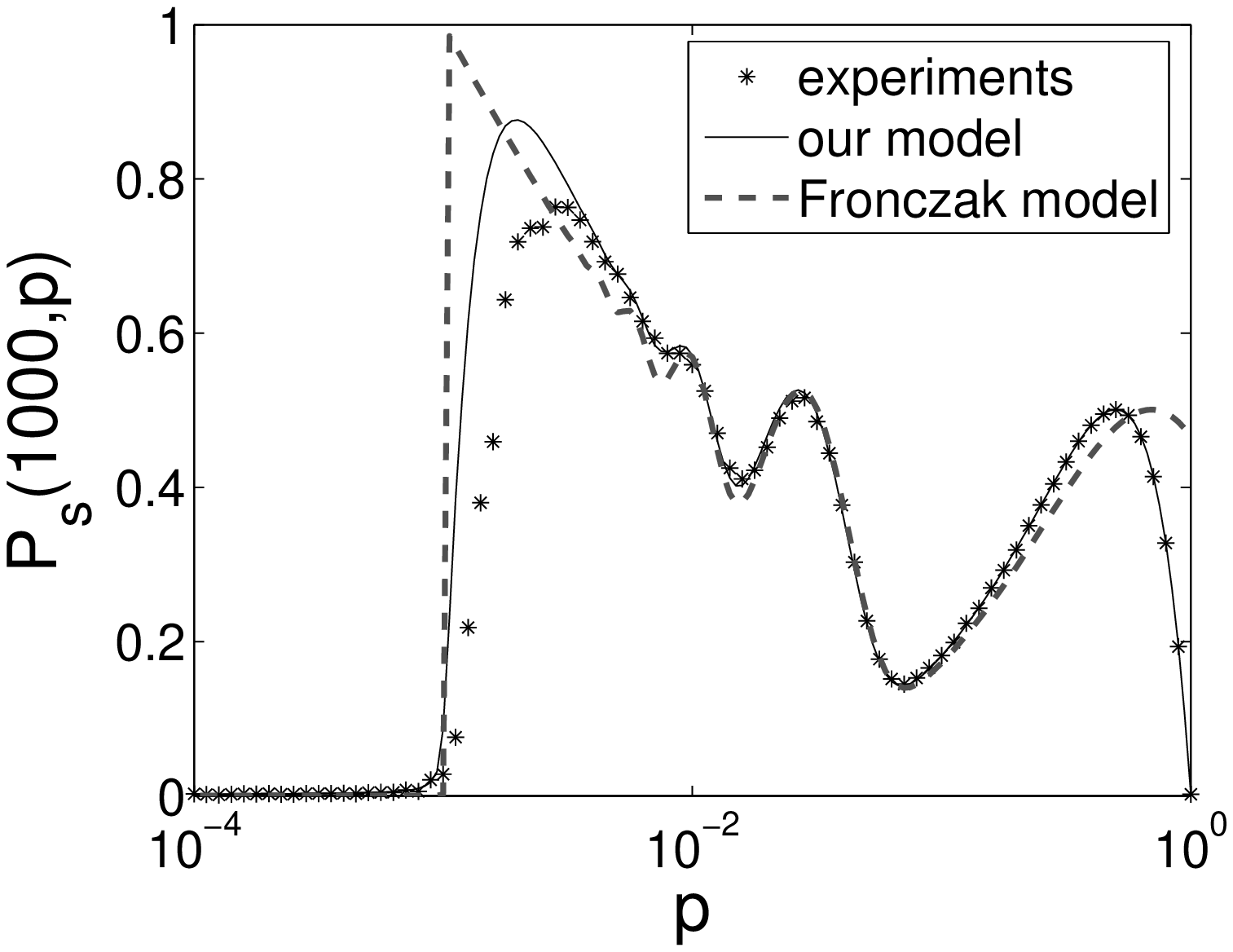}\\(a)\\\\
\includegraphics[scale = 0.5]{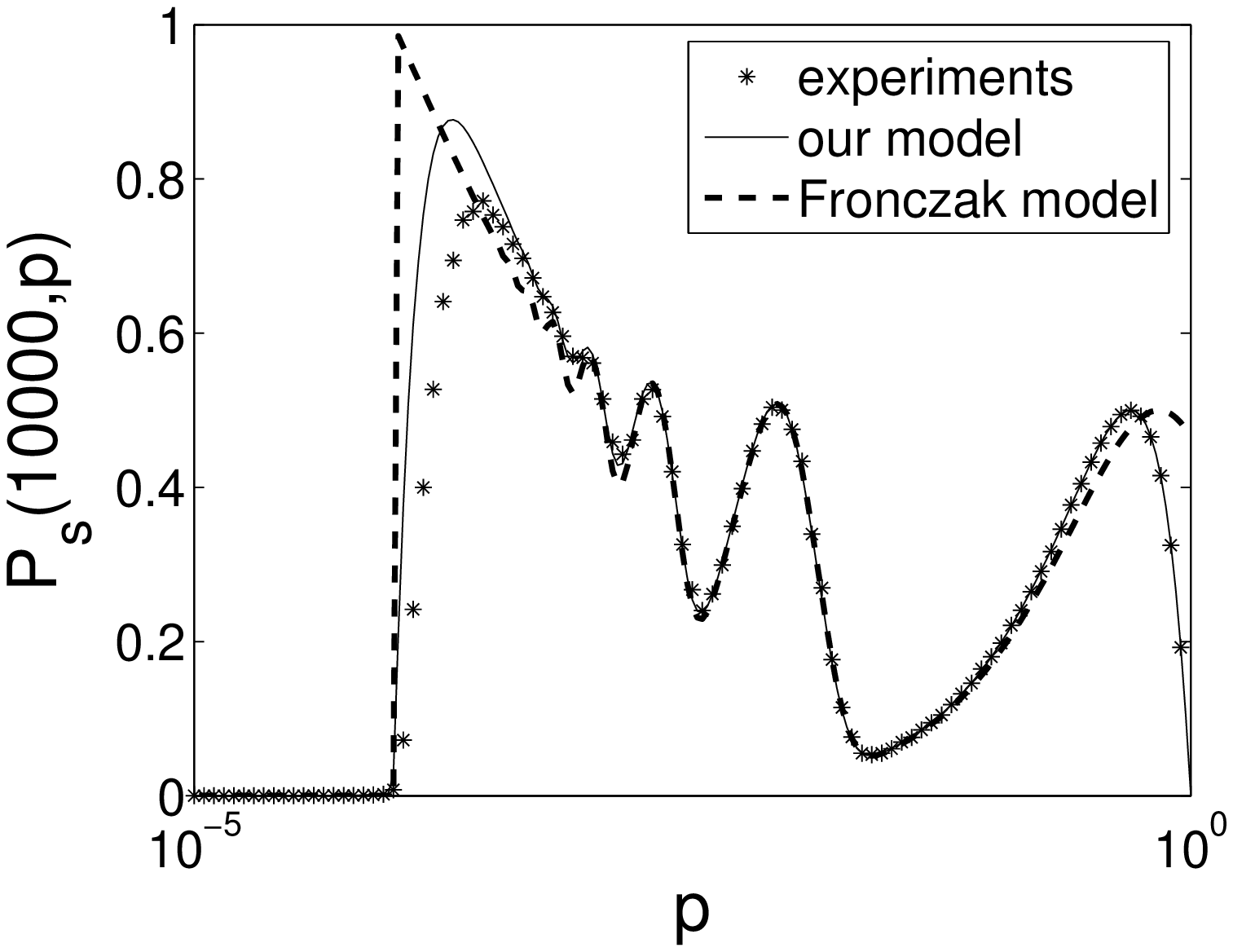}\\(b)
\end{tabular}

\caption{Comparison of the evolution of $P_s(n,p)$ {with $n=1000$
(a) and $n=10000$ (b)} according to numerical experiments
\cite{jlg}, to our model, and to Fronczak et al.'s model. }
\label{fig-camel_comp}
\end{figure}

We close this section by explaining how the distance distribution
can be used to compute the proportion $P_s$ of edges belonging to
shortest paths starting at the source node. As explained in the
introduction, the edges that do not belong to any shortest path
are those connecting nodes that are at the same distance from the
source, in addition to all edges that are not in the same
connected component as the source. Since the expectation of the
number of nodes at distance $d$ from the source is equal to
$nf_d$, the expected number of edges connecting these nodes is
roughly equal to $\frac{1}{2}p \prt{nf_d}^2$.  Taking
$\frac{1}{2}pn^2$ as the total number of edges, we obtain the
following expression for the proportion of edges that lie on a
shortest path in an \erdren graph, which we denote by $P_s(n,p)$
in the {sequel:} {
\begin{equation}\label{eq:P_C_def} P_s(n,p) = 1 -
\frac{\sum_{d=1}^n p \prt{nf_d}^2}{n^2 p}= 1-\sum_{d=1}^n {f_d^2}.
\end{equation}
Note that this expression implicitly handles the edges that are
not in the same connected component as the source if we take
$f_{n}=F_{n-1}\approx F_{\infty}$.  Indeed, this quantity
represents those nodes that are not connected to the source, as
they are at a distance larger than $n-1$.} The evolutions with $p$
of $P_s$ using the two models presented above are represented in
FIG. \ref{fig-camel_comp} for $n=1000$ and $n=10000$. One can see
that our results match the experiments very well except when $np$
is larger than but close to one, which is the range of parameters
for which our model has already been shown to be less accurate.
{Moreover, the range of values of $np$ for which our model is less
accurate appears not to grow with $n$.}

\section{Analysis of the curve}\label{section-analysis}

\begin{figure}
\centering
\begin{tabular}{c}
\includegraphics[scale = .5]{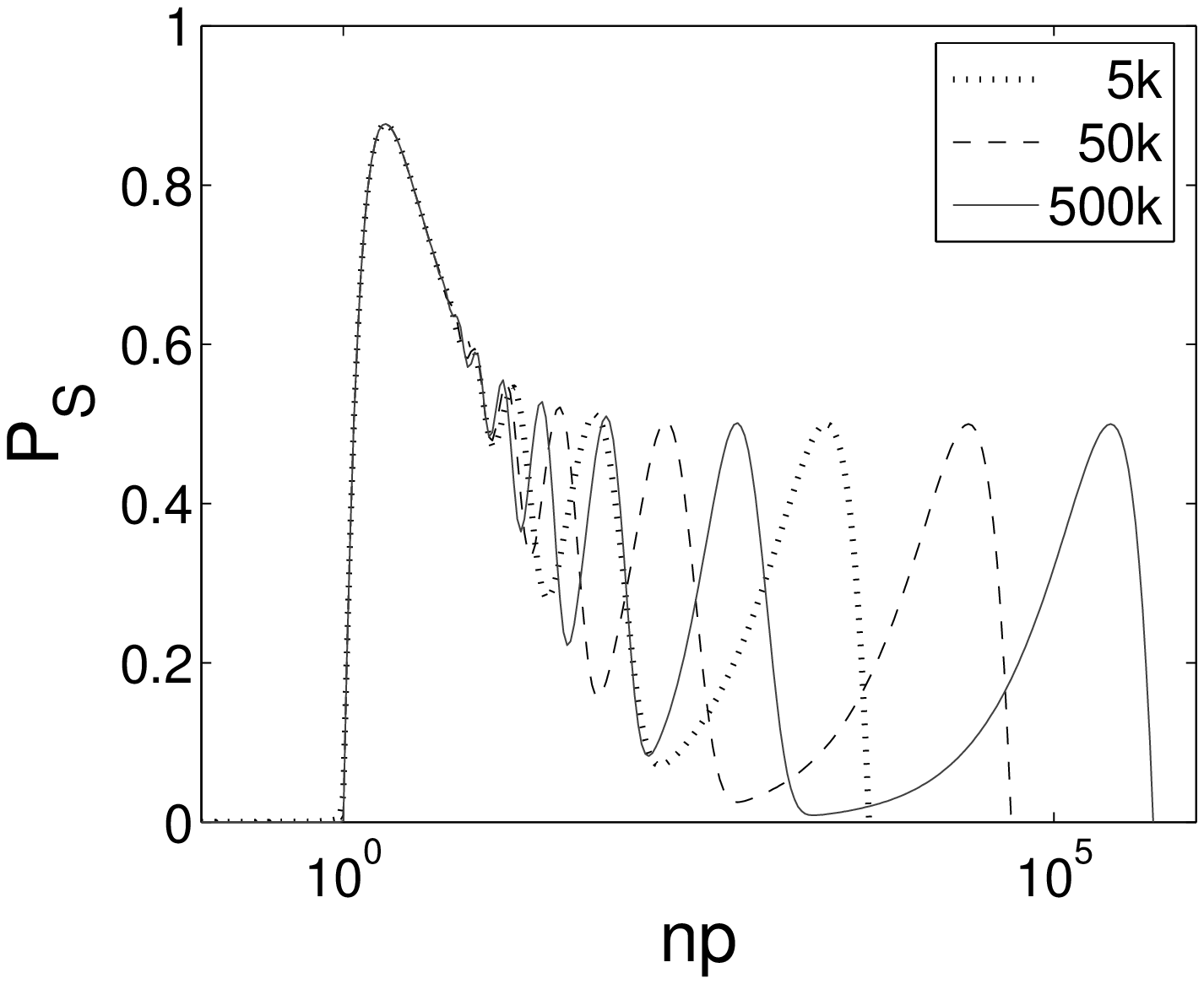}\\
(a)\\
\includegraphics[scale = .5]{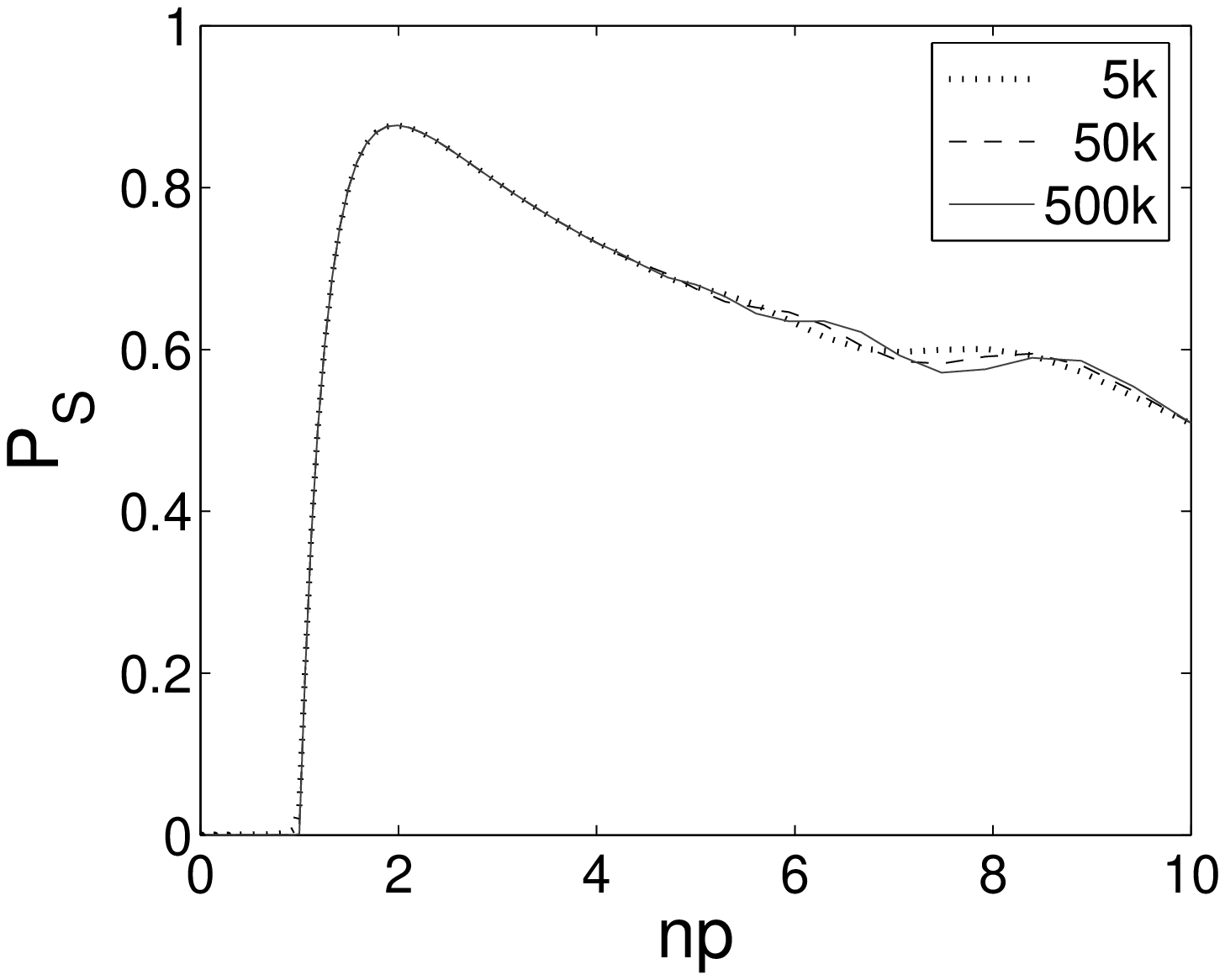}\\
(b)
\end{tabular}
 \caption{Evolution of $P_s(n,p)$ with $np$ for
different values of $n$. All curves present a sharp increase
between $np=1$ and $np=2$, and a global maximum in {$np\simeq 2$}.
For larger values, the curves present several oscillations, with
local maxima tending to 0.5. (b) is a zoomed-in linear-scale
version of (a).}\label{fig-compar_np}
\end{figure}

In this section we analyze the function $P_s(n,p)$ generated with
our model and (\ref{eq:P_C_def}).  {We show the appearance of a
sort of phase transition: for some particular values, a weak
variation of the probability $p$ may cause abrupt changes in the
proportion of discovered edges with the ASP model, and affect
dramatically the properties of the graph.} We give analytical
formulas for the asymptotic behavior in
several phases.\\
We begin by analyzing the first transition, starting from small
values of $p$. It is well known \cite{rand-graphs} that in an
\erdren graph, a giant component emerges when $p$ becomes larger
than $1/n$. If the average degree $np$ is sufficiently small, the
graph is not connected and the only edges that the observer can
see are in the (small) connected component of the source. This
quantity is negligible in view of the total number of edges, and
so the function is approximately zero. Note however that such
graphs do not contain many cycles, so that most paths starting
from the source are shortest paths. Therefore, the observer
discovers approximately all edges in its connected component. When
$p$ grows the size of the connected components increases, so that
more and more edges are discovered. Now when $np\approx 1$, the
giant component emerges very quickly, and the source is in this
component with a large probability. Since most of the edges are
also in this component, the proportion of discovered edges
increases rapidly with $np$. Simultaneously with the apparition of
a giant component, there also appears a non negligible number of
cycles in the graphs, so that not all edges lie on shortest paths
anymore. As a result of these two conflicting phenomena a (global)
optimum is reached for $np\approx 2$. Experimentally our model
gives an optimum that seems to lie exactly at $np=2,$ but we have
not been able to prove this, nor to express analytically the
values of $P_s(n,p)$ around $np\approx 2.$ However, {experiments}
seem to indicate that in this range of parameters $P_s(n,p)$ only
depends on $np.$  All this can be seen in FIG.
\ref{fig-compar_np}, for different values of $n$. When $np$
becomes larger, one can see that $P_s$ does not only depend on
$np$, and presents an oscillatory behavior. In particular, the
successive values of the maxima seem to tend to $\frac{1}{2}$. We
explain
this phenomenon in the sequel.\\

\begin{figure}
\centering
\includegraphics[scale =0.5]{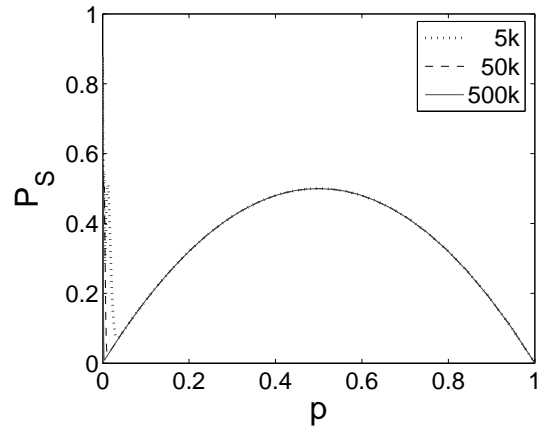}
\caption{Evolution of $P_s(n,p)$ with $p$ for different values of
$n$. On any interval $[\epsilon,1]$, $P_s$ tends to the parabola
$2p(1-p)$ when $n$ increases.} \label{fig-parabole}
\end{figure}

\begin{figure}
\centering
\begin{tabular}{c}
\includegraphics[scale = .5]{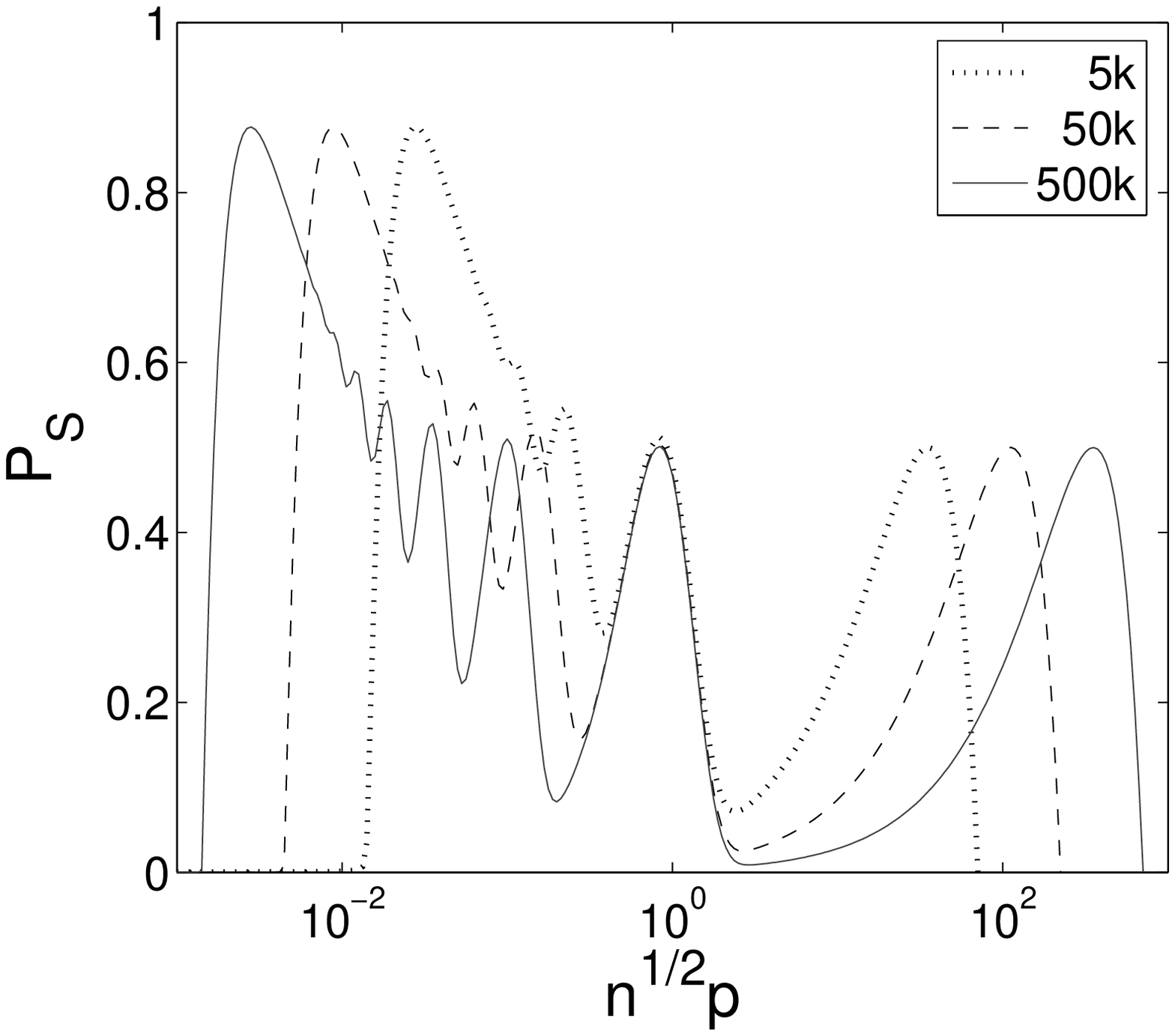}\\
(a)\\
\includegraphics[scale = .5]{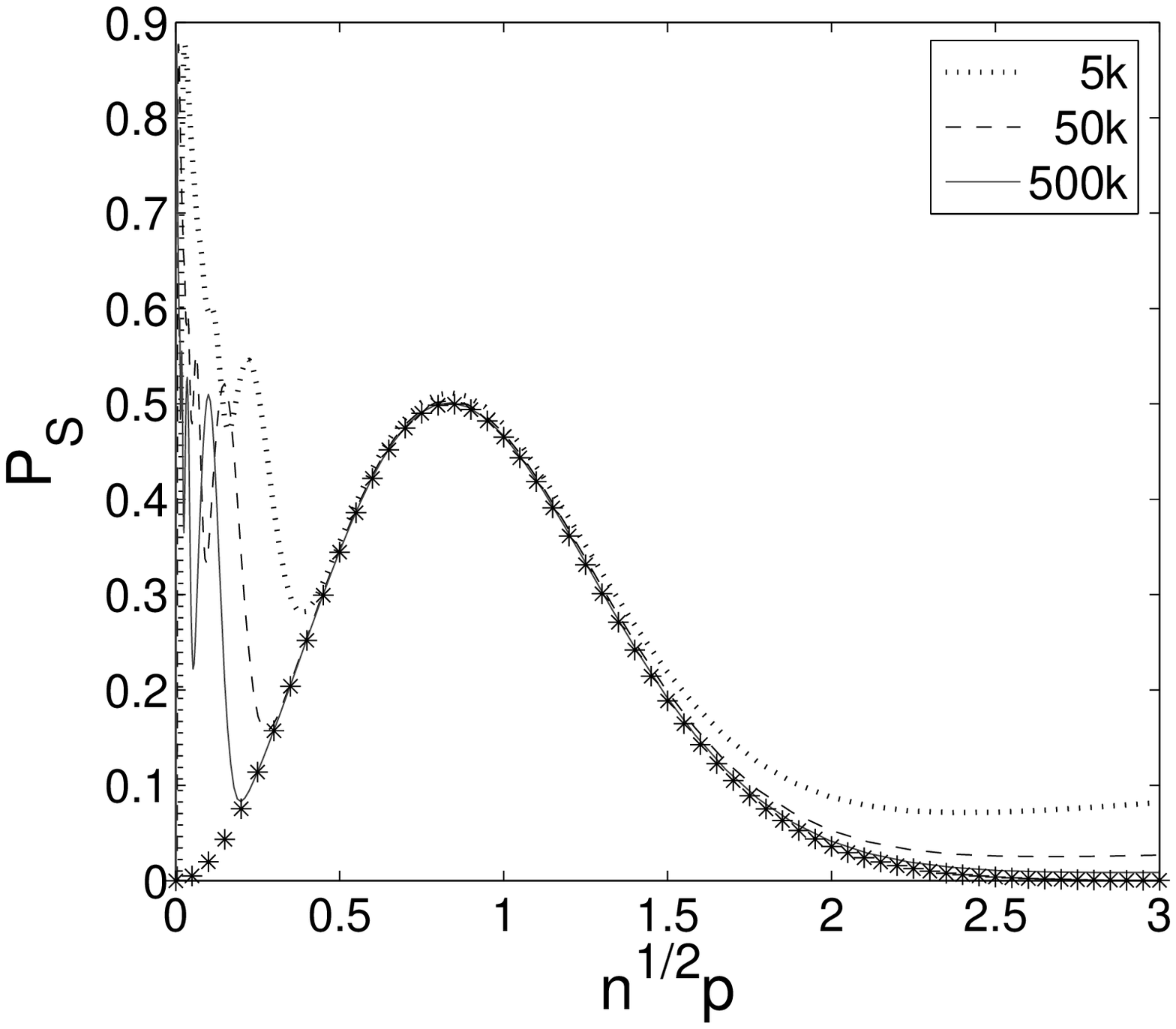}\\
(b)
\end{tabular}
 \caption{Evolution with $n^{\frac{1}{2}}p$ of $P_s(n,p)$ for
different values of $n$. (b) is a zoomed-in linear-scale version
of (a). Asymptotically, $P_c$ behaves as
$2e^{-np^2}\prt{1-e^{-np^2}}$ which is represented by \quotes{$*$}
in (b).}\label{fig-compar_n1/2p}
\end{figure}

As can be observed in FIG. \ref{fig-parabole}, the shape of
$P_s(p)$ tends to the parabola $2p(1-p)$ on any interval
$[\epsilon,1]$ when $n$ increases (note that the $x$-axis is in
linear scale). This fact can be proved theoretically, based on our
model of evolution of $F_d$. In the sequel, for the sake of
clarity in our analysis, we modify (\ref{eq-rec-app}), and study
the slightly different one:
\begin{equation}\label{eq:simplifiedrecur}
F_d(n,p) = (1-p)^{n(1-F_{d-1}(n,p))}.
\end{equation}
This new approximation is justified by the fact that we will
consider asymptotic behaviors for $n\rightarrow \infty$. Moreover,
the results that we derive can be obtained without making this
approximation. Observe that $F_1(n,p)= 1-p$, so that $F_2(n,p) =
(1-p)^{np}$. When $n$ grows $F_0 = 1-\frac{1}{n}\rightarrow 1$,
and if $p$ is bounded from below by an arbitrary positive constant
$\epsilon,$ $F_2(n,p) = (1-p)^{np}$ tends uniformly to 0. As a
consequence the probability $f(d)$ for a node to be at a distance
$d$ from the source tends uniformly to 0 for all $d$ except for
$d=1,2$, for which $f_1 = F_0(n,p)-F_1(n,p) \rightarrow p$ and
$f_2 = F_1(n,p)-F_2(n,p) \rightarrow 1-p$. It follows then from
(\ref{eq:P_C_def}) that
\begin{equation*}
P_s(n,p) \rightarrow 1 - p^2 - (1-p)^2 = 2p(1-p),
\end{equation*}
so that asymptotically, the last maximum of $P_s$ is $\frac{1}{2}$
and is reached at $p=\frac{1}{2}$. The asymptotic parabolic
character of $P_s$ is thus here due to the fact that almost all
nodes tend to be at a distance either 1 or 2 from the source when
$n$ grows and $p$ is sufficiently large, as can for example be
observed in FIG. \ref{fig-comp_F(d)}(d).\\
We now analyze the oscillating behavior between the first and last
maximum. One can see in FIG. \ref{fig-compar_n1/2p} that around
the second rightmost maximum, $P_s$ only depends on
$n^{\frac{1}{2}}p,$ and that $P_s$ asymptotically behaves as
\begin{equation}\label{eq:d=2}
P_s\simeq
2e^{-(n^{\frac{1}{2}}p)^2}\prt{1-e^{-(n^{\frac{1}{2}}p)^2}}
\end{equation}
around this maximum.  The maximum therefore tends to $\frac{1}{2}$
when $n\rightarrow \infty$ and is attained for
$(n^{\frac{1}{2}}p)^2 = \log 2$.  To explain (\ref{eq:d=2}), we
show in the appendix that similarly as above, all nodes are
asymptotically at distance either $2$ or $3$ when $n\rightarrow
\infty$ with $\epsilon <np^2 <R,$ where $\epsilon, R$ are
arbitrarily positive constants.
 As in the case of the parabola, this together with (\ref{eq:P_C_def})
implies that $P_s$ then asymptotically behaves as $2(1-F_2)F_2.$
We also show that $F_2(n,p)$ tends to $e^{-np^2},$ which implies
(\ref{eq:d=2}).\\

\begin{figure}
\centering
\begin{tabular}{c}
\includegraphics[scale = .5]{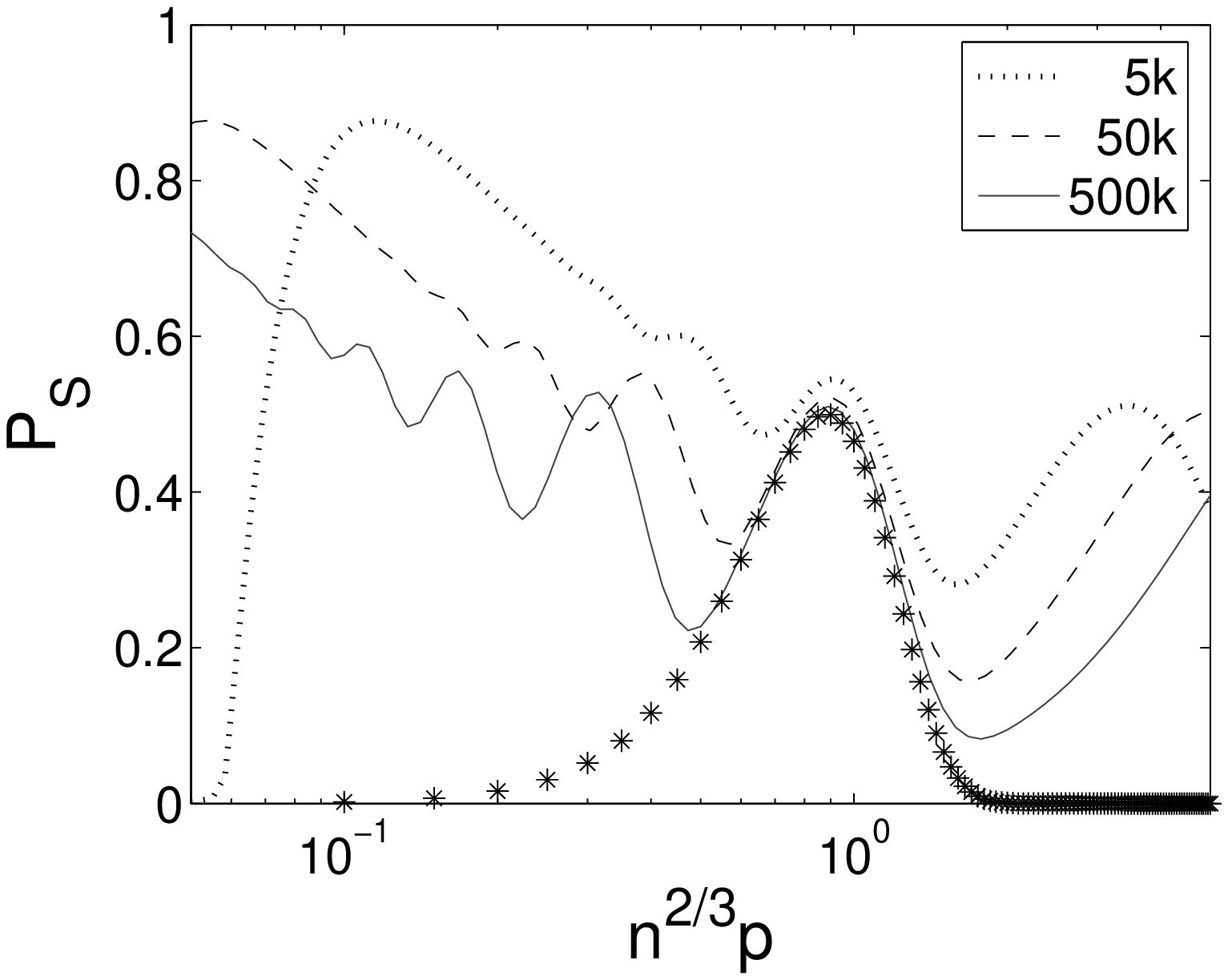}\\
(a)\\
\includegraphics[scale = .5]{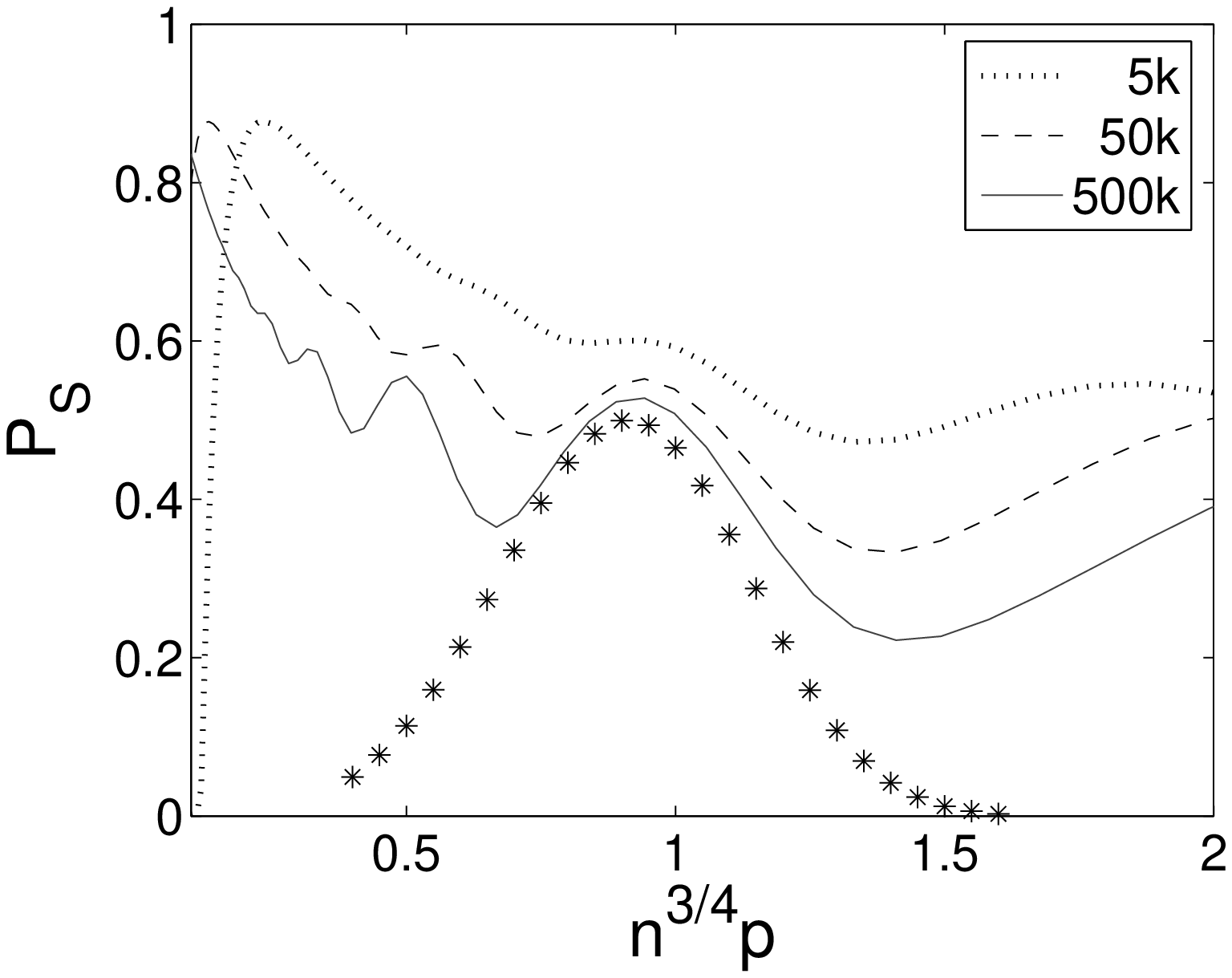}\\
(b)
\end{tabular}
\caption{Evolution of $P_s(n,p)$ with $n^{\frac{2}{3}}p$ (a) and
$n^{\frac{3}{4}}p$ (b), for different values of $n$.
Asymptotically, local maxima $\frac{1}{2}$ appear for
$n^{\frac{2}{3}}p=\sqrt[3]{\log 2}$ and
$n^{\frac{3}{4}}p=\sqrt[4]{\log 2}$.  The \quotes{$*$} represent
the theoretical asymptotic behavior.} \label{fig-comparnk-1kp}
\end{figure}

Actually the previous relations can be generalized inductively: we
prove in the appendix that when $n\rightarrow \infty$ with
$\epsilon<n^{d-1}p^d<R,$ $F_d$ converges uniformly to
$e^{-n^{d-1}p^d},$ while all $F_{d'}$ with $d'<d$ converge
uniformly to 1 and all others to 0.  This means that in this range
of parameters, and when $n$ tends to infinity, almost all nodes
are at distance $d$ or $d+1$ from the source. It follows then from
(\ref{eq:P_C_def}) that
\begin{equation*}
\lim_{\epsilon<n^{d-1}p^d<R}P_s(n,p) = 2 \prt{1-e^{-n^{d-1}p^d}
}e^{-n^{d-1}p^d},
\end{equation*}
which, as for $d=1,2$, is a parabolic curve with respect to $F_d$.
This parabolic curve attains its maximum $\frac{1}{2}$ when
$e^{-n^{d-1}p^d}= \frac{1}{2}$. So, when $n\rightarrow \infty$,
$P_s$ contains an unbounded number of oscillations and local
maxima with asymptotic values $\frac{1}{2}$, and these maxima are
attained when $n^{d-1}p^d = \log 2$ for each $d>1$ as can be seen
on some additional examples in FIG. \ref{fig-comparnk-1kp}.
Experimentally, all local maxima but the first global one can be
explained in that way. Between two maxima, there is a zone where
asymptotically $F_d \simeq 1$ and $F_{d+1}\simeq 0$, so that
almost all nodes are at distance $d+1$ from the source, and $P_s
\simeq 1-1^2 =0$. Such behavior is obtained when $n\rightarrow
\infty$ with either large values of $n^{d-1}p^d$ but still
$\epsilon<n^{d-1}p^d<R$, or small values of $ n^{d}p^{d+1}$ but
still $\epsilon<n^{d}p^{d+1}<R$. One can indeed see in FIG.
\ref{fig-compar_n1/2p} and \ref{fig-comparnk-1kp} for example that
the values of the local minima decrease significantly when $n$
increases.  Let us mention that an explanation of the oscillatory
behavior based on the fact that almost all nodes are at distance
$d$ or $d+1$ from the source had been suggested without proof in
\cite{jlg}.\\

\section{Conclusions and future work}\label{section-conclusions}

The goal of this paper {was} twofold: First, we have proposed a
simple model for the computation of the inter-vertex distance
distribution in a random graph, via a recurrence equation for the
probability for two randomly chosen nodes to be at distance more
than $d$.  Contrary to the model of Fronczak et al., our
recurrence equation is not explicitly solvable, but it is more
accurate. It has to be noted that for the range of parameters
corresponding to the oscillating behavior analyzed at the end of
Section \ref{section-analysis}, the two models are equally valid,
and that the analysis that we have made for such values could also
be made using Fronczak et al.'s model. Let us add that the ideas
behind the derivation of the formula remain valid for more general
graphs such as random graphs with hidden variables. In the
particular case of \erdren graphs, these ideas lead to a simple
recurrence equation, allowing to compute explicit values
numerically, and to prove the asymptotic behavior of the curve
experimentally obtained in \cite{jlg}. Nevertheless, a further
analysis for more general graphs could be interesting.\\

Second, following numerical simulations in previous works
motivated by practical graph exploration questions \cite{jlg}, we
have analyzed the proportion of edges connecting nodes that are
equidistant from a certain source node in random graphs. The
evolution of this quantity with the parameter $p$ exhibits an
intriguing oscillating behavior, which we have been able to
explain and reproduce with a great accuracy using our model. We
have also characterized precisely the (infinite number of)
transitions for this quantity, and the analytical
evolution with $p$ in the different phases.\\

\begin{bf}
Acknowledgment
\end{bf}

The authors wish to thank Renaud Lambiotte for his useful advice.


 \appendix

\section{Expression of the asymptotic behavior}\label{section-ap}

In this appendix, we provide an analytical expression for $F_d$
when $n$ tends to infinity with $\epsilon<n^{d-1}p^d <R$, and we
show that in this range of parameters, almost all nodes are at
distance $d$ or $d+1.$  Suppose first that $n\rightarrow \infty$
with $0<np^2 < R$ for an arbitrary $R$. Then $p \rightarrow 0$ so
that $F_1 = (1-p) \rightarrow 1$ uniformly with $np^2$. From our
recurrence formula (\ref{eq:simplifiedrecur}), we have
\begin{equation*}
F_2(n,p) = (1-p)^{np} = \prt{(1-p)^\frac{1}{p}}^{np^2},
\end{equation*}
which, together with the classical relation $\lim_{p\rightarrow
0}(1-p)^{\frac{1}{p}}=e^{-1},$ implies that
\begin{equation}\label{F2}
\lim_{np^2<R}F_2(n,p) = e^{-np^2}
\end{equation}
holds uniformly for $0<np^2<R$. We now show that
\begin{equation*}
\lim_{\epsilon<np^2<R}F_3(n,p)=\lim_{\epsilon<np^2<R}
(1-p)^{n(1-F_2)} = 0
\end{equation*}
for any two arbitrary constants $\epsilon$ and $R.$ This implies
that almost all nodes are at distance $2$ or $3$ from the source.
It follows from (\ref{F2}) that $1-F_2$ is uniformly bounded from
below by a positive constant when $n\rightarrow\infty$ with
$\epsilon<np^2 <R$, so that we just need to prove the uniform
decay of $(1-p)^n$. The latter expression can be rewritten as
\begin{equation}\label{eq:1-p^nbase}
(1-p)^n=\prt{\prt{1-\frac{np^2}{(np)}}^{(np)}} ^\frac{1}{p}.
\end{equation}
Since $np\rightarrow \infty$ when $n\rightarrow\infty$ with
$\epsilon<np^2 <R$, there uniformly holds
\begin{equation*}
e^{-R }\leq\lim_{\epsilon<np^2 <R}
\prt{1-\frac{np^2}{(np)}}^{np}\leq e^{-\epsilon}.
\end{equation*}
And since $\frac{1}{p}\rightarrow \infty,$ it follows from
(\ref{eq:1-p^nbase}) that
\begin{equation*}
\lim_{\epsilon<np^2 <R} (1-p)^n = 0,
\end{equation*}
which implies the desired result.\\

There remains to prove our assertions about the asymptotic
behavior of $F_d$ for any $d>2$. We first prove by induction that
the two following relations hold uniformly for $n^{d-1}p^d <R$
where $R$ is any arbitrary positive constant.
\begin{equation}
\label{eq:limd-1} \lim_{n^{d-1}p^d <R} F_{d-1}(n,p) =1,
\end{equation}
\begin{equation}
\label{eq:limd}
 \lim_{n^{d-1}p^d <R} F_{d}(n,p) =e^{-n^{d-1}p^d}.
\end{equation}
These relations hold for $d=2$ as shown above. Let us now assume
that they hold for a certain $d-1$ and prove that they then hold
for $d$. Observe first that when $n\rightarrow \infty$ with
$n^{d-1}p^d < R$, $n^{d-2}p^{d-1}$ tends uniformly to 0 and is
bounded. It follows then from the induction hypothesis that
\begin{equation*}
F_{d-1}(n,p) \rightarrow e^{-n^{d-2}p^{d-1}} \rightarrow 1 -
n^{d-2}p^{d-1}
\end{equation*}
uniformly when $n\rightarrow \infty, n^{d-1}p^d <R.$  So Equation
(\ref{eq:limd-1}) is proved. By our recurrence relation
(\ref{eq:simplifiedrecur}), $F_d(n,p) = \prt{1-p}^{n(1-F_{d-1})}$.
Therefore, there holds
\begin{equation*}
\lim_{n^{d-1}p^d <R} F_d(n,p) = \prt{1-p}^{(np)^{d-1}} =
\prt{\prt{1-p}^{\frac{1}{p}}}^{n^{d-1}p^d}.
\end{equation*}
Since $n^{d-1}p^d$ is bounded, and since $p$ tends thus uniformly
to 0 when $n\rightarrow \infty$, the last equation becomes
\begin{equation*}
\lim_{n^{d-1}p^d <R} F_d(n,p) = e^{-n^{d-1}p^d}
\end{equation*}
uniformly for $n^{d-1}p^d\in(0,R)$, which proves (\ref{eq:limd}).\\

Using the results above we now prove that for any $d>2$, the
following holds uniformly
\begin{equation}\label{eq:fk+1=0}
\lim_{\epsilon<n^{d-1}p^d <R} F_{d+1}(n,p) = 0,
\end{equation}
where $\epsilon$ and $R$ are two arbitrary positive constants. By
(\ref{eq:simplifiedrecur}), we have
\begin{equation*}
F_{d+1}(n,p)=\prt{1-p}^{n(1-F_d)}.
\end{equation*}
It follows from the results above that $1-F_d$ is uniformly
bounded from below by a positive constant when
$n\rightarrow\infty$ with $\epsilon<n^{d-1}p^d <R$, so that we
just need to prove the uniform decay of $(1-p)^n$. The latter
expression can be rewritten as
\begin{equation}\label{eq:1-p^n}
(1-p)^n=\prt{\prt{1-\frac{n^{d-1}p^d}{(np)^{d-1}}}^{(np)^{d-1}}}
^\frac{1}{n^{d-2}p^{d-1}}.
\end{equation}
Since $(np)^{d-1}\rightarrow \infty$ when $n\rightarrow\infty$
with $\epsilon<n^{d-1}p^d <R$, there uniformly holds
\begin{equation*}
e^{-R }\leq\lim_{\epsilon<n^{d-1}p^d <R}
\prt{1-\frac{n^{d-1}p^d}{(np)^{d-1}}}^{(np)^{d-1}}\leq
e^{-\epsilon}.
\end{equation*}
And since $n^{d-2}p^{d-1}\rightarrow 0$ when $n\rightarrow\infty$
with $\epsilon<n^{d-1}p^d <R$, it follows from (\ref{eq:1-p^n})
that
\begin{equation*}
\lim_{\epsilon<n^{d-1}p^d <R} (1-p)^n = 0,
\end{equation*}
which implies the desired result (\ref{eq:fk+1=0}).

\end{document}